\documentclass[12pt]{article}

\usepackage{amsmath}
\usepackage{amsfonts}
\usepackage{latexsym}
\usepackage{a4wide}
\usepackage{euscript}
\usepackage{exscale}

\newcommand{\M}{\mbox{${\mathcal M}$}}
\renewcommand{\H}{\mbox{${\mathcal H}$}}

\renewcommand{\O}{\mbox{${\mathcal O}$}}
\newcommand{\A}{\mbox{${\mathcal A}$}}
\newcommand{\EuS}{\mbox{${\EuScript S}_\varepsilon$}}
\newcommand{\skalar}[2]{\mbox{$\left\langle #1 \left| #2 \right. 
\right\rangle$}}         
\newcommand{\Cnull}[1]{\mbox{$C_{0,\mathbb R}^\infty(#1)$}}
\newtheorem{theo}{Theorem}
\newtheorem{defi}{Definition}

\begin{document}

\title{A KMS-like state of Hadamard type on Robertson-Walker spacetimes 
  and its time evolution}
\author{Mathias Trucks\thanks{\hspace{0.1cm}  
    e-mail: trucks@physik.tu-berlin.de}\\
       \small Institut f\"ur Theoretische Physik,\\ 
       \small Technische Universit\"at Berlin\\
       \small Hardenbergstra{\ss}e 36, 10623 Berlin, Germany\\[0.3cm]}
\date{}
\maketitle

\begin{abstract}
  In this work we define a new state on the Weyl algebra of the free
  massive scalar Klein-Gordon field on a Robertson-Walker spacetime
  and prove that it is a Hadamard state. The state is supposed to
  approximate a thermal equilibrium state on a Robertson-Walker
  spacetime and we call it an adiabatic KMS state.  This opens the
  possibility to do quantum statistical mechanics on Robertson-Walker
  spacetimes in the algebraic framework and the analysis of the free
  Bose gas on Robertson-Walker spacetimes. The state reduces to an
  adiabatic vacuum state if the temperature is zero and it reduces to
  the usual KMS state if the scaling factor in the metric of the
  Robertson-Walker spacetime is constant.
  
  In the second part of our work we discuss the time evolution of
  adiabatic KMS states. The time evolution is described in terms of
  semigroups.  We prove the existence of a propagator on the classical
  phase space. This defines a time evolution on the one-particle
  Hilbert space. We use this time evolution to analyze the evolution
  of the two-point function of the KMS state. The inverse temperature
  change is proportional to the scale factor in the metric of the
  Robertson-Walker spacetime, as one expects for a relativistic Bose
  gas.
\end{abstract}

\section{Introduction}

Beyond the standard model of cosmology the inflationary scenario,
involving phase transitions and symmetry breaking, has been
intensively discussed during the last years. For an investigation of
such phenomena it is necessary to describe the thermal behavior of
quantum fields and states. In this work we start a discussion of
quantum statistical mechanics on Robertson-Walker spacetimes in the
framework of algebraic quantum field theory. An analysis of the free
Bose gas on Robertson-Walker spacetimes could serve as a model for
more complicated quantum field theories. The system may show a phase
transition, namely Bose-Einstein condensation.

The algebraic framework of quantum field theory started with the work
of Haag and Kastler \cite{HaagKastler}; for an overview and basic
results see the book of Haag \cite{Haag}. Dimock \cite{Dimock80}
generalized the axioms to globally hyperbolic spacetimes. Basic
object is a net of $C^*$-algebras arising from the assignment of a
$C^*$-algebra \A(\O) to each open, relatively compact subset \O{} of a
manifold \M{}.  The algebra \A(\O) is the algebra of local
observables, i.e.~the observables that can be measured in the region
\O. The quasilocal algebra $\A$ is defined as the inductive limit of
the \A(\O), i.e.~$\A =\overline{\cup_{\mathcal{O} \in \mathcal{M}}
  \A(\O)}$, where the bar denotes the norm closure and the union runs
through all relatively compact open sets.

A state is a positive normalized linear functional on \A.  One of the
major problems in algebraic quantum field theory on curved spacetimes
is to pick out physically relevant states among all positive
normalized linear forms. In quantum field theory on Minkowski
spacetime there is the Poincar\'{e} group as the symmetry group and
the spectrum condition which sort out the Minkowski vacuum.  There is
no symmetry group on a generic curved spacetime and therefore no way
to distinguish a vacuum-like state.

In quantum field theory on curved spacetimes the class of Hadamard
states is believed to be a class of physically relevant states. We
mention two reasons supporting this opinion:
\begin{enumerate}
\item The class of Hadamard states allows the renormalization of the
  energy-momentum tensor $T_{\mu\nu}$. Quantum field theory on curved
  spacetimes is a semi-classical theory, where matter fields are
  quantized but not the metric, so one has to deal with the
  semiclassical Einstein equation
  \[ G_{\mu\nu} = 8\pi\langle T_{\mu\nu}\rangle_\omega     \,.  \]
  The energy-momentum tensor contains products of fields and their
  derivatives at one point. To the right-hand side has to be given
  sense in a state $\omega$ by a renormalization procedure. For
  Hadamard states the renormalization by a point-splitting procedure is
  possible (see the book of Wald \cite[Chap.~4.6]{Wald} and references
  therein). 
\item Haag et al.~\cite{HaagNarn} formulated the ``principle of local
  definiteness'', which contains requirements to physically relevant
  states, namely local quasi-equivalence, local primarity and local
  definiteness.  It was shown by Verch \cite{Verch94} that all Hadamard
  states on a globally hyperbolic spacetime are locally quasi-equivalent.
  For ultrastatic spacetimes he showed that the local von Neumann
  algebras arising from Hadamard states are factors (of type
  III$_1$), i.e.~they are local primary and he also showed local
  definiteness. Recently he strengthened these results to arbitrary globally
  hyperbolic spacetimes \cite{Verch97}. 
\end{enumerate}
For these reasons it is reasonable to consider Hadamard states as
good candidates for physically relevant states.

There are not many explicitly known Hadamard states although Junker
\cite[Chap.~3.7]{Junker} has given an explicit construction of
Hadamard states. As explicitly known Hadamard states we mention the
ground state on ultrastatic spacetimes, KMS states on ultrastatic
spacetimes with compact spacelike Cauchy surfaces, adiabatic vacua on
Robertson-Walker spacetimes and the Hartle-Hawking state on extended
Schwarzschild spacetime.

The class of adiabatic vacuum states is a class of Hadamard states,
defined on the Weyl algebra of the free massive Klein-Gordon field on
Robertson-Walker spacetimes, which approximates a vacuum state.  On
the other hand it is possible to define KMS states, i.e.~thermal
equilibrium states, on ultrastatic spacetimes, e.g.~the Einstein
static universe, in a usual way. In this paper we combine these
definitions to define a state which approximates a thermal equilibrium
state on Robertson-Walker spacetimes, an adiabatic KMS state. We will
prove that this state is a Hadamard state, i.e.~belongs to the class of
physically relevant states.  We generalize our definition by
introducing a chemical potential $\mu$. The state is a Hadamard state,
if $\mu < m$, where $m$ is the mass parameter in the Klein-Gordon
equation. We remark that Bose-Einstein condensation sets in, if the
value of the chemical potential reaches the value of the mass
parameter.

We also show that an adiabatic KMS state satisfies the KMS condition
with respect to an automorphism group. This automorphism group does
not generate the time translations of the system. This possibility was
already mentioned in the fundamental work on KMS states by Haag,
Hugenholtz and Winnink \cite{HHW}.

The general idea of adiabatic KMS states and vacua is to maintain as
many properties as possible of KMS states and ground states in the
ultrastatic case. But it is known that a ``naive'' generalization does
not lead to Hadamard states \cite[Chap.~3.6]{Junker}. The ``positive
frequencies'' cannot be fixed on a Cauchy surface, but must be
determined dynamically off the Cauchy surface.  We think of an
adiabatic KMS state as a state which approximates a thermal
equilibrium state in the sense that switching off the expansion of the
Robertson-Walker spacetime would lead to a thermal equilibrium state
of inverse temperature $\beta$ for $t \to \infty$.  This can be seen
with the methods of the second part of the work.

In the second part of the work we analyze the evolution of an
adiabatic KMS state. We introduce new coordinates, so that the
Klein-Gordon equation can be written as a first order system having
only off-diagonal entries. This defines a semigroup for fixed $t$. We
prove the existence of a propagator on the classical phase space.  By
a natural generalization of the notion of a one-particle Hilbert space
structure we are able to define the time evolution on the one-particle
Hilbert space. It is given by a unitary propagator. This unitary
operator is not identical to the unitary operator coming from the
group of automorphisms with respect to which an adiabatic KMS state
satisfies the KMS condition. It can be seen that an adiabatic vacuum
state is invariant under this time evolution in a certain sense.  For
an adiabatic KMS state the inverse temperature is proportional to the
scale factor $R$ in the Robertson-Walker metric, as one expects for a
relativistic Bose gas. Similarly one could show for a non-relativistic
Bose gas the inverse temperature change to be proportional to $R^2$.

The work is organized as follows. In the next section the notion of
adiabatic vacua is reviewed. After some preliminary remarks on
one-particle Hilbert space structures and the definition of KMS states
we give the definition of adiabatic KMS states in section 3.  In
section 4 we give a precise definition of Hadamard states and prove
that adiabatic KMS states are of this kind. A section on the KMS
condition follows. In the following section we describe the time
evolution in terms of semigroups. The existence of a propagator is
proved. The time evolution on the one-particle Hilbert space is
described in section~\ref{sec:time_evolution}. In the last section we
compute the evolution of adiabatic KMS states. The necessary results
on pseudodifferential operators and wave-front sets are summarized in
the appendix.

\section{Adiabatic vacuum states}
\label{sec:avs}

In this section we briefly summarize the definition of adiabatic
vacua. For a more detailed discussion see \cite{Junker,LuedersRoberts,
Trucks96}. Readers only interested in the definition of an adiabatic
KMS state can skip over this section. Adiabatic vacua were originally
introduced by Parker \cite{Parker69I}.

We consider the Klein-Gordon equation 
\[
   (\Box_g - m^2)\varphi = 0 \; ,
\]
on a Lorentz manifold $(\M,g)$ topologically of the form $\M = I
\times \EuS$, $I \subset {\mathbb R}$, where $\varepsilon=1,0,-1$
corresponds to the spherical, the flat and the hyperbolic case,
respectively and $g$ is a Robertson-Walker metric
\begin{equation}\label{metric}
  g = -dt^2 + R(t)^2[d\theta_1^2 + \Sigma^2_\varepsilon(d\theta_2^2
  + \sin^2\theta_2 d\phi^2)] = -dt^2 + h_{ij}(\EuS)dx^idx^j
  \, , \quad i,j = 1,2,3,
\end{equation}
with $\Sigma_1 = \sin\theta_1, \Sigma_0 = \theta_1, \Sigma_{-1} =
\sinh\theta_1$ and $R(t) > 0$,  

We construct the Weyl algebra $CCR(D,\sigma)$ associated with the
space of classical solutions of the Klein-Gordon operator on
Robertson-Walker spacetimes, where $D$ is a real vector space and
$\sigma$ a symplectic form on $D$ (see e.g.~\cite[8.2]{BW}).
 
As the real symplectic vector space we consider the space of real
Cauchy data on a Cauchy surface $D := \Cnull{\EuS} \oplus
\Cnull{\EuS}$. Let $E$ be the uniquely existing causal propagator of
the Klein-Gordon operator (see Dimock \cite{Dimock80}). For a function
$f \in \Cnull{\M}$ the mapping to $D$ is given by $\rho_0 Ef \oplus
\rho_1 Ef$, where $\rho_0: \Cnull{\M} \to \Cnull{\EuS}$ is the
restriction operator to the Cauchy surface and $\rho_1: \Cnull{\M} \to
\Cnull{\EuS}$ is the forward normal derivative on the Cauchy surface.
The Weyl algebra $CCR(D,\sigma)$ is the algebra generated by the
elements $W(F) \neq 0, F \in D$, obeying the Weyl form of the
canonical commutation relations
\[
   W(F)W(G) = e^{-i\sigma(F,G)/2}W(F+G)\, , \quad F,G \in D,
\]
and with the property $W(F)^* = W(-F)$, see e.g.~\cite[8.2]{BW}. 
We define the symplectic form by 
\[
   \sigma(F,G) = \int_{\EuS} (f_1g_2 - f_2g_1) \,d\mu(\EuS),
   \quad F=f_1\oplus f_2, \quad  G=g_1\oplus g_2 \, ,
\]
where $d\mu(\EuS) = \sqrt{\det h_{ij}(\EuS)}d^3x$ is the invariant
measure on the spaces $\EuS$.  The algebras of local observables are
\[
   \A(\O) = C^*(W(F), \; F
   \in D, \; \mbox{supp}\;(F) \subset \O \subset \M  \; ) \; ,
\]
where we mean the $C^*$-algebra generated by these elements. 

The Klein-Gordon operator on $(\M,g)$ has the form
\[
    \Box_g - m^2 = -\partial_t^2 - 3 H(t)
    \partial_ t + R^{-2}(t) \Delta_\varepsilon - m^2  \, ,
\]
where $ \Delta_{\varepsilon}$ is the Laplace operator on the
respective spatial parts, $H(t) = \dot{R}(t)/R(t)$ and $\partial_t
\equiv \partial/\partial t$, The eigenvectors and eigenvalues of the
Laplace operator are explicitly known in these three cases. This fact
allows us to separate the time-dependent part of the equation.
Denoting the eigenvectors by $Y_{\vec{k}}(\underline{x})$ and the
eigenvalues of the Laplace operator by $-E(k)$, i.e. $\Delta
Y_{\vec{k}}(\underline{x}) = -E(k)Y_{\vec{k}}(\underline{x})$, we can
express the elements of the set of Cauchy data $D$ on a Cauchy surface
with the uniform notation
\begin{eqnarray*} 
  F &=& \binom{f_1}{f_2} =
  \int d\vec{k} \binom{c(\vec{k})}{\hat{c}(\vec{k})}
  Y_{\vec{k}}(\underline{x}) \, , \\ G &=& \binom{g_1}{g_2} = \int d\vec{k}
  \binom{b(\vec{k})}{\hat{b}(\vec{k})} Y_{\vec{k}}(\underline{x}) \, ,
  \quad F,G \in D \, ,
  \nonumber 
\end{eqnarray*} 
where the integral reduces to a sum in the spherical case (see Junker
\cite[3.5]{Junker} for details). The main result concerning the
structure of Fock states is summarized in the following
\begin{theo} 
  The Fock states for the Klein-Gordon field
  on a Robertson-Walker spacetime are given by a two-point function of the form
  \begin{eqnarray}\label{zpfunk}
    \skalar{F}{G}_S &=& \int d\vec{k} [
    \overline{c(\vec{k})}b(\vec{k}) S_{00}(k) + 
    \overline{c(\vec{k})}\hat{b}(\vec{k}) S_{01}(k) \nonumber \\ & & {}+
    b(\vec{k})\overline{\hat{c}(\vec{k})} S_{10}(k) +
    \overline{\hat{c}(\vec{k})}\hat{b}(\vec{k}) S_{11}(k) ] \,.
  \end{eqnarray}
  The entries of the matrix $S$ can be expressed in the form
  \begin{eqnarray}\label{matrix} 
    S_{00}(k) &=& |p(k)|^2 \, , \qquad
    S_{11}(k) = R^6|q(k)|^2 \, , \nonumber \\ S_{01}(k) &=& -R^3q(k)
    \overline{p(k)} \, , \qquad S_{10} = \overline{S}_{01} \,, 
  \end{eqnarray} 
  where $p$ and $q$ are polynomially bounded measurable
  functions satisfying
  \begin{equation}\label{bedi} 
    \overline{q(k)}p(k) - \overline{p(k)}q(k) = i \,.  
  \end{equation} 
  Conversely every pair of polynomially bounded measurable functions
  satisfying equation (\ref{bedi}) yields via (\ref{matrix}) and
  (\ref{zpfunk}) the two-point function of a Fock state.
\end{theo} 
For the proof see L\"uders and Roberts \cite[Thm.~2.3]{LuedersRoberts}.

We consider the time-dependence of the Klein-Gordon equation
\begin{equation}\label{t2kg}
   \left( d_t^2 + 3 H(t) \,
     d_t + m^2 + R^{-2}(t)E(k) \right) T_k(t) = 0 , \quad \forall k \,.
\end{equation}
This equation can be solved explicitly only in exceptional cases. In
the general case, one tries to solve it by an iteration procedure.
For finding the iteration, we consider
\[              
        T_{k}(t) = [2R^3(t)\Omega_k(t)]^{-1/2} \exp 
                \left(i \int_{t_0}^t \Omega_k(t') \, dt' \right) \, ,
                \quad \forall k \, ,
\]
where the functions $\Omega_k$ have to be determined. Inserting this
ansatz in equation (\ref{t2kg}) we find that the functions $\Omega_k$
have to satisfy
\begin{equation}\label{iteration} 
  \Omega_k^2 = \omega_k^2 - \frac{3}{4}
  \left( \frac{\dot{R}}{R} \right)^2 - \frac{3}{2}
  \frac{{\ddot R}}{R} + \frac{3}{4} 
  \left( \frac{\dot{\Omega}_k}{\Omega_k} \right)^2
  -\frac{1}{2} \frac{\ddot{\Omega}_k}{\Omega_k} \, ,
\end{equation}
where $\omega_k^2 = E(k)/R^2 + m^2$. With
\[
  (\Omega_k^{(0)})^2 := \omega_k^2 = E(k)/R^2 + m^2  
\]
the iteration is given by
\[
  (\Omega_k^{(n+1)})^2 = \omega_k^2 - \frac{3}{4}
  \left( \frac{\dot{R}}{R} \right)^2 - \frac{3}{2}
  \frac{{\ddot R}}{R} + \frac{3}{4} 
  \left( \frac{\dot{\Omega}_k^{(n)}}{\Omega_k^{(n)}} 
  \right)^2 -\frac{1}{2} \frac{\ddot{\Omega}_k^{(n)}}
  {\Omega_k^{(n)}}     \,.                 
\]
The functions $T_k(t)$ and $\dot{T}_k(t)$ are related to the 
functions $q(k)$ and $p(k)$, which constitute the matrix $S$
by equation (\ref{matrix}). On a Cauchy surface at time $t$ 
these relations are
\begin{equation}\label{adia1}
  T_k(t) = q(k) \, , \quad \dot{T}_k(t) = R^{-3}(t)p(k)  \,.
\end{equation}
An adiabatic vacuum state will now be defined by initial values
at time $t$:
\begin{defi}
  For $t_0,t \in {\mathbb R}$, let
  \begin{equation*}
    W_k^{(n)}(t) := [2R^3(t)\Omega_k^{(n)}(t)]^{-1/2} 
    \exp \left(i \int_{t_0}^t \Omega_k^{(n)}(t') \, dt' \right) \,.
  \end{equation*}
  An adiabatic vacuum state of order $n$ is a Fock state, obtained via
  equations (\ref{matrix}) and (\ref{adia1}), where the initial values
  at time $t$ for equation (\ref{t2kg}) can be expressed by
  \[
    T_{k}(t) =      W_k^{(n)}(t) \, , \qquad 
    \dot{T}_{k}(t) = \dot{W}_k^{(n)}(t) \,.
  \]
\end{defi}
For later purposes we notice that for an adiabatic vacuum state of zeroth 
order we have
\begin{equation}
  S_{11}(k) = R^6|q(k)|^2 = R^6|T_k(t)|^2 = \frac{R^3}{2\omega_k}\,.
\end{equation}

\section{Adiabatic KMS states}

We review the definitions of ground and KMS states on ultrastatic spacetimes by
their one-particle Hilbert space structure. Then we describe an
adiabatic vacuum state (of zeroth order) in a similar way. A short
computation shows the coincidence of this definition with the one given by
L\"uders and Roberts \cite{LuedersRoberts}. Adiabatic KMS states are
defined by imitating the connection of KMS states on ultrastatic spacetimes
with the ground state on the same spacetime.

\subsection{One-particle Hilbert space structures}

Let $\omega_S$ be a state on the Weyl algebra $CCR(D,\sigma)$. A {\bf
  one-particle Hilbert space structure} is a real-linear map $K:D \to
\H$, $\H$ a Hilbert space, satisfying
\begin{enumerate}
  \label{one_particle}
  \item $KD + iKD$ is dense in $\H$,
  \item $[S(F,G) + i\sigma(F,G)]/2 = \skalar{KF}{KG} \, , \quad F,G \in D$,
\end{enumerate}
where $S(\cdot,\cdot)$ is a real scalar product on $D$ and
$\omega_S(W(F)) = \exp (-S(F,F)/4)$ is the generating functional of
the state $\omega_S$.  Usually it is also required that the map $K$
intertwines the time evolutions on the phase space $D$ and the Hilbert
space \H. We come back to this point in
section~\ref{sec:time_evolution}.

\subsubsection{One-particle structure $K$ for a ground state on 
  ultrastatic spacetimes} 
\label{groundstate}
For a ground state on an ultrastatic spacetime, the one-particle
Hilbert space structure is given by
\[
K:D \to \H = L^2_{\mathbb C}({\EuScript S},d\mu), \quad f_1 \oplus f_2 \mapsto
2^{-1/2}(A^{1/4}f_1 + i A^{-1/4}f_2) \, , \quad A := m^2 - \Delta \, ,
\]
where $\Delta$ is the Laplacian on the Cauchy surface $\EuScript S$.

\subsubsection{One-particle structure $K^\beta$ for a KMS state on 
  ultrastatic spacetimes} 

For a KMS state of inverse temperature $\beta$, the one-particle
Hilbert space structure is defined by doubling the Hilbert space:
\[
K^\beta : D \to \H \oplus \H\, , \qquad F \mapsto(\cosh Z^\beta) KF
\oplus  C(\sinh Z^\beta) KF \, ,
\]
where $Z^\beta$ is implicitly defined by $\tanh Z^\beta = \exp(-\beta
A^{1/2})$, i.e.
\[
  \cosh^2 Z^\beta = [1 - \exp(-\beta A^{1/2})]^{-1} \, , \quad \sinh^2
  Z^\beta = [1 - \exp(-\beta A^{1/2})]^{-1} \exp(-\beta A^{1/2})\, ,
\]
$C$ is a conjugation and $K$ is the map defined in 
subsubsection~\ref{groundstate}.

\subsubsection{One-particle structure $K^a_t$ for an adiabatic vacuum state}

An adiabatic vacuum state (of zeroth order) can also be described by a
one-particle Hilbert space structure, as we will show below.  The
mapping which defines an adiabatic vacuum state is given by
\begin{eqnarray*}
  K^a_t : D &\to& \H = L^2_{\mathbb C}(\EuS,d\mu)\, , \qquad f_1 \oplus
  f_2 \mapsto B_1(t) f_1 + i B_2(t) f_2 \, , \\ B_1(t) &=&
  2^{-1/2}A^{1/4}(t)\{1 + iH(t)[A^{-1/2}(t) +
  m^2 A^{-3/2}(t)/2]\} \, , \\ B_2(t) &=&
  2^{-1/2}A^{-1/4}(t) \, ,
\end{eqnarray*}
where $A(t) = m^2 - \Delta/R^2(t)$.  This one-particle Hilbert space
structure leads to a two-point function
\begin{eqnarray*}
  \lefteqn{\skalar{K^a_t(f_1\oplus f_2)}{K^a_t(g_1\oplus g_2)} = 
  \skalar{B_1 f_1 + i B_2 f_2}{B_1 g_1 + i B_2 g_2} }         \\
  &=& \skalar{\dbinom{f_1}{f_2}}{\begin{pmatrix}B_1^* B_1 & iB_1^* B_2 \\
      -iB_2^* B_1 & B_2^* B_2\end{pmatrix} \dbinom{g_1}{g_2}} \,.
\end{eqnarray*}
We show this expression to be equivalent to the two-point function of
an adiabatic vacuum state of zeroth order as defined by L\"uders and
Roberts \cite{LuedersRoberts}. For example for the fourth entry we
have
\begin{eqnarray*}
  2\skalar{f_2}{B_2^*B_2g_2} &=& \skalar{f_2}{A^{-1/2}g_2} = \\
  &=& \skalar{\int d\vec{k}\;\hat{c}(\vec{k}) Y_{\vec{k}}(\underline{x})}
  {A^{-1/2}\int d\vec{k'}\;\hat{b}(\vec{k'}) Y_{\vec{k'}}(\underline{x})} = \\
  &=& \int d\vec{k}\;\overline{\hat{c}(\vec{k})} 
  \int d\vec{k'}\;\hat{b}(\vec{k'})\omega_{k'}^{-1}
  \skalar{Y_{\vec{k}}(\underline{x})}{Y_{\vec{k'}}(\underline{x})} = \\
  &=& \int d\vec{k}\;\overline{\hat{c}(\vec{k})} 
  \int d\vec{k'}\;\hat{b}(\vec{k'})\omega_{k'}^{-1}
  R^3\delta (\vec{k}-\vec{k'}) = \\
  &=& R^3 \int d\vec{k}\;\overline{\hat{c}(\vec{k})} 
  \omega_{k}^{-1}\,\hat{b}\,(\vec{k}) \,.
\end{eqnarray*}
Comparing this with equation (\ref{zpfunk}) gives $S_{11}(k) = 
R^3/(2\omega_k)$, which is the desired result of equation (\ref{zpfunk}).

\subsection{Definition of an adiabatic KMS state}

If we look at the definition of KMS states on an ultrastatic
spacetime, we find that it is connected with the ground state on this
spacetime. In a similar way we connect an adiabatic KMS state with an
adiabatic vacuum state:
\begin{defi}
  \label{adiakms}
  We define an \textbf{adiabatic KMS state} by a one-particle Hilbert
  space structure $K^{a\beta}$ given by
  \begin{eqnarray*}
    K^{a\beta}_t : D \;&\to& \;{\mathcal H} \oplus {\mathcal H}  \, ,\\
    F \; &\mapsto& \; (\cosh Z^\beta) K^a_t F \oplus C(\sinh Z^\beta)
    K^a_t F \, ,
  \end{eqnarray*}
  where $K^a_t$ is the one-particle structure of an adiabatic vacuum
  state, $C$ is a conjugation and  $\tanh Z^\beta = \exp(-\beta
  A^{1/2}(t))$. 
\end{defi}

This definition leads to the following two-point function of an
adiabatic KMS state 
\begin{eqnarray*} 
  \lefteqn{\skalar{K^{a\beta}(f_1\oplus f_2)}{K^{a\beta}(g_1\oplus
      g_2)} = } \\ &=& \skalar{B_1 f_1 + i B_2 f_2}{\cosh^2
    Z^\beta (B_1 g_1 + i B_2 g_2)} + \skalar{B_1^* f_1 - i B_2^*
      f_2}{\sinh^2Z^\beta (B_1^* g_1 - i B_2^* g_2)} \, ,
\end{eqnarray*}
so that the two-point distribution $\Lambda$ has the form (we suppress
the $t$-dependence of $A$): 
\begin{eqnarray}
\label{zweipunkt}
\lefteqn{\Lambda(f,g) = }  \\ &=&
\frac{1}{2}\skalar{[A^{1/2}+iH(1+\frac{m^2}{2}A^{-1})
  +i\partial_t]Ef} {A^{-1/2}\cosh^2Z^\beta[A^{1/2}
  +iH(1+\frac{m^2}{2}A^{-1})+i\partial_t]Eg} \nonumber \\
&+&
 \frac{1}{2}\skalar{[A^{1/2}-iH(1+\frac{m^2}{2}A^{-1})
  -i\partial_t]Ef}{A^{-1/2}\sinh^2Z^\beta[A^{1/2}-
  iH(1+\frac{m^2}{2}A^{-1})-
  i\partial_t]Eg} \, ,\nonumber 
\end{eqnarray}
for $f,g \in \Cnull{\M}$ and $E$ is the causal propagator (see
section~\ref{sec:avs}).  In section~\ref{sec:hadamardproof} we will
prove that this two-point distribution is the two-point distribution
of a Hadamard state.

\section{Hadamard property of an adiabatic KMS state}

In the next section we give a precise definition of a Hadamard state
and prove in the following subsection that an adiabatic KMS state is a
Hadamard state. The necessary results on pseudodifferential operators
and wave-front sets are summarized in the appendix.

\subsection{Hadamard states}

Since the work of Radzikowski \cite{Radzikowski} it is known that
Hadamard states can be characterized by the wave-front set of its
two-point distribution.  Earlier definitions required a specific form
of the two-point distribution (see Kay and Wald \cite{KayWald}). The
characterization of a Hadamard state by its wave-front set is easier
to handle and offers new possibilities to prove the Hadamard property
of a state, but it requires some knowledge of pseudodifferential
operators (PDO's) and wave-front sets of distributions. We refer to the
appendix for notation and some results used below.

\begin{defi}
  A quasifree state of a Klein-Gordon quantum field on a globally
  hyperbolic spacetime is a {\bfseries Hadamard state} iff the
  wave-front set of its two-point distribution $\Lambda$ is of the form:
  \begin{equation}
  \label{Hadamard}
    WF(\Lambda) = \{ (x_1,\xi_1;x_2,-\xi_2) \in T^*(\M \times
    \M)\setminus\{0\} \; | \; (x_1,\xi_1) \sim (x_2,\xi_2), \xi_1^0 \geq 0
    \} \, ,
  \end{equation}
  where the notation $(x_1,\xi_1) \sim (x_2,\xi_2)$ means that $x_1$ and
  $x_2$ can be joined by a null geodesic $\gamma$ and $\xi_1$ is
  tangent to $\gamma$ in $x_1$ and $\xi_2$ is the parallel transport
  of $\xi_1$ along $\gamma$ in $x_2$.
\end{defi}
The proof that a state is a Hadamard state requires only the analysis
of the wave-front set of its two-point distribution. We will do this in
the next section for an adiabatic KMS state.

\subsection{An adiabatic KMS state is a Hadamard state}
\label{sec:hadamardproof}

In the proof that an adiabatic KMS state is a Hadamard state we use
the following theorems due to Junker \cite[Thm.3.11 and 3.12]{Junker}.
\begin{theo}\label{one-particle}
  Let $(\M,g)$ be a globally hyberbolic spacetime with Cauchy surface
  $\EuScript S$ and $(D,\sigma)$ be the phase space of initial data on
  $\EuScript S$ of the Klein-Gordon field. \\
  Let $B, I, S, C$ be operators on $L^2_{\mathbb{R}}({\EuScript
    S},d\mu)$, such that $I$ is symmetric, $B$ is selfadjoint,
  positive and invertible and $C^*C - S^*S = 1$.\\
  Then, with ${\mathcal H} = L^2_{\mathbb{C}}({\EuScript S},d\mu)$,
  \begin{eqnarray*}
    K: D &\to& \tilde{\H} = \H \oplus \H \, , \\
    (f_1,f_2) &\mapsto& C(2B)^{-1/2}[(B + iI)f_1 + if_2]\oplus 
    S(2B)^{-1/2}[(B - iI)f_1 - if_2] \, ,
  \end{eqnarray*}
  is a one-particle Hilbert space structure.
\end{theo}
For a proof see Junker \cite[Thm.~3.11]{Junker} (where different
conventions are used).

Under the assumption that the metric has the form of equation
(\ref{metric}), the two-point distribution $\Lambda$ resulting from
this one-particle structure is given by
\begin{eqnarray}\label{twopoint}
  \Lambda(f,g) &=& \frac{1}{2}\skalar{(B + iI + i\partial_t)Ef}
  {B^{-1/2}C^*CB^{-1/2}((B + iI + i\partial_t)Eg} + \nonumber \\ 
   & &{}  + \frac{1}{2}\skalar{(B - iI - i\partial_t)Ef}
  {B^{-1/2}S^*SB^{-1/2}(B - iI - i\partial_t)Eg} \, , 
\end{eqnarray}
where $f,g \in \Cnull{\M}$ and $E$ is the causal propagator (see
section~\ref{sec:avs}).

Now let $(\M,g)$ be a globally hyperbolic spacetime, foliated in a
neighborhood of $\EuScript S$ into $(-T,T) \times {\EuScript S}$ with
${\EuScript S}_t := \{t\} \times {\EuScript S}$ and ${\EuScript S}_0 =
\EuScript S$ and $g$ of the form given in equation (\ref{metric}).
\begin{theo}\label{junker}
  Let $B(t),I(t),S(t),C(t)$ be PDO's on ${\EuScript S}_t, t
  \in (-T,T)$, satisfying the properties stated in 
  theorem~\ref{one-particle}, such that $B$ is elliptic, $S \in
  OPS^{-\infty}$, and such that there exists a PDO $Q$ on $\M$ with the
  property $Q(B + iI +i\partial_t) = \Box_g - m^2$ which possesses a
  principal symbol $q$ with
  \begin{equation*}
    q^{-1}(0)\setminus \{0\} \subset \{(x,\xi) \in T^*(\M)\, | \, \xi^0
    \geq 0 \} \,.
  \end{equation*}
  Then the quasifree state given by the one-particle Hilbert space
  structure of theorem~\ref{one-particle} is a Hadamard state,
  i.e.~the wave-front set of the corresponding two-point distribution
  $\Lambda$ has the form of equation (\ref{twopoint}).
\end{theo}
For a proof see Junker \cite[Thm.~3.12]{Junker}.

This theorem will be used in the proof of the following theorem. The
proof for closed Robertson-Walker spacetimes is in fact a
generalization of Junker's proof \cite[Chap.~3.4]{Junker} that a KMS
state on an ultrastatic spacetime with compact spacelike Cauchy
surface is a Hadamard state.
\begin{theo}
  An adiabatic KMS state on the Weyl algebra of the free massive Klein-Gordon
  field on Robertson-Walker spacetimes as defined in definition~\ref{adiakms} 
  is a Hadamard state.
\end{theo}
Proof: We have to show that the wave-front set of the two-point
distribution (\ref{zweipunkt}) has the form of equation (\ref{Hadamard}). \\
For $F=f_1 \oplus f_2 \in D$, we have
\[
K^a_t F = (2A^{1/2})^{-1/2}\{[A^{1/2}+iH(1+m^2A^{-1}/2)]f_1 + i f_2\} \,.
\]
We identify the operator $B$ in theorem~\ref{one-particle}
respectively theorem~\ref{junker} with
$A^{1/2}=(m^2-\Delta/R^2)^{1/2}$, which is an elliptic, selfadjoint,
positive PDO (of order 1).  Furthermore the operator $I$ is identified
with $H(1+m^2A^{-1}/2)$, which is a symmetric PDO. The operators $S$
respectively $C$ are identified with $\sinh Z^\beta$
respectively $\cosh Z^\beta$, so that $C^*C - S^*S = 1$.

For closed Robertson-Walker spacetimes we proceed as follows: Since
\begin{equation*}
S = \exp(-\beta A^{1/2}/2)(1-\exp(-\beta A^{1/2}))^{-1/2}
\end{equation*}
and $A^{1/2}$ has the properties of theorem~\ref{funktionalkalkuel} in
the appendix with the real-valued principal symbol given by $a(x,\xi)
= (h^{ij} \xi_i \xi_j)^{1/2}$, we can apply this theorem, to conclude
that $S$ is a PDO with principal symbol
\[ 
  p(a(x,\xi)) = \frac{\exp(-\beta
  a(x,\xi)/2)}{(1-\exp(-\beta a(x,\xi)))^{1/2}} \,.
\] 
This principal symbol falls off faster than any inverse power of
$\xi$, so that $S \in OPS^{-\infty}$ and this also means that $C \in
OPS^0$.

For flat Robertson-Walker spacetimes we show directly that the involved 
operators are PDO's:
\begin{eqnarray*}
  (\cosh Z^\beta f)(t,\underline{x}) 
  &=& \left( 1 - \exp(-\beta A^{1/2}) )^{-1/2} f \right)(t,\underline{x}) = \\
  &=& (2\pi)^{-3/2} \int ( 1 - \exp(-\beta\omega_k) )^{-1/2} 
  \tilde{f}(t,\vec{k}) Y_{\vec{k}}(\underline{x}) \; d\vec{k} \, ,
\end{eqnarray*}
which is a PDO of order zero, because 
\[
  a(k) = (1-\exp(-\beta\sqrt{k^2 + m^2}))^{-1/2}
\]
is a symbol of order zero. Furthermore
\[
  (\sinh Z^\beta f)(t,\underline{x}) 
  = (2\pi)^{-3/2} \int  \frac{\exp(-\beta\omega_k/2)}
  {(1-\exp(-\beta\omega_k))^{1/2}}
  \tilde{f}(t,\vec{k}) Y_{\vec{k}}(\underline{x}) \; d\vec{k} \, ,
\]
is a PDO of order $-\infty$, because
\[
  a(k) = \frac{\exp(-\beta\sqrt{k^2 + m^2}/2)}
  {(1-\exp(-\beta\sqrt{k^2 + m^2}))^{1/2}}
\]
and the derivatives of $a(k)$ tend to zero as $k \to \infty$.  For
hyperbolic Robertson-Walker spacetimes one has to express the
eigenfunctions of the Laplace operator in terms of
$\exp(i\underline{x}\cdot\underline{\xi})$ in the same way as it is
done in \cite[Proof of Lemma 3.26]{Junker}, to see that the operators
are PDO's of the desired type.

The operator $Q$ is given by
\[
  Q = 3iH/2 - A^{-1/2}\partial_t A^{1/2} -  A^{1/2} +
  i\partial_t \,.
\]
This can be verified with the help of equation (\ref{iteration}).
\hfill $\Box$\\[0.5cm]

\subsection{Introduction of a chemical potential}

We generalize the definition to the case of a non-vanishing chemical
potential $\mu$. This can be done by changing the definition of $\tanh
Z^\beta$: Let 
\begin{equation}\label{chemical}
\tanh Z^\beta = \exp (-\beta h(\mu)) \, , \qquad h(\mu) = A^{1/2}(t) -
\mu \,.
\end{equation}
The operator $h(\mu)$ is selfadjoint on $\mbox{dom}(A^{1/2})$,
positive if $\mu < m$ and elliptic. So we can generalize the proof in
the case of a closed Robertson-Walker spacetimes to the case of a
non-vanishing chemical potential $\mu$ under the restriction $\mu <
m$. In the case of non-closed Robertson-Walker spacetimes we can again
directly compute it.
\begin{theo}
  An adiabatic KMS state on the Weyl algebra of the free massive
  Klein-Gordon field on Robertson-Walker spacetimes as defined
  in definition~\ref{adiakms} generalized by equation (\ref{chemical}) is
  a Hadamard state if $\mu < m$.
\end{theo}

\section{On the KMS condition}
\label{sec:kms}

In this section we show that an adiabatic KMS state fulfills the KMS
condition with respect to an automorphism group $\alpha_s$. It is not
the automorphism group that generates the time translations of the
system. It was already remarked in the fundamental paper on the KMS
condition by Haag, Hugenholtz and Winnink \cite{HHW}, that such a
situation can occur. This means the system were in equilibrium if time
evolution were given by the automorphism group $\alpha_s$, i.e.~if
$h(s) = (m^2-\Delta/R^2(s))^{1/2}$ were independent of $s$.

A KMS state can be defined in the following way (see
Kay and Wald \cite{KayWald}).
\begin{defi}\label{kms}
  Let $\alpha_s$ be an automorphism group on a $C^*$-algebra
  ${\mathcal A}$ and $\omega$ an $\alpha_s$-invariant state. $\omega$
  is a KMS state at inverse temperature $\beta$ if its GNS triple
  $({\mathcal F},\pi_\beta, \Omega_\beta)$ satisfies the following
  properties:
  \begin{enumerate}
  \item The unique unitary group $U(s): {\mathcal F} \to {\mathcal F}$
    which implements $\alpha_s$ and leaves $\Omega_\beta$ invariant,
    is strongly continuous, so that $U(s) = \exp(-iHs)$ for some
    selfadjoint operator $H$.
  \item \label{ii} $\pi_\beta({\mathcal A})\Omega_\beta$ is contained in the
    domain of $\exp(-\beta H/2)$.
  \item There exists a complex conjugation $J$ on ${\mathcal F}$
    satisfying 
    \[
    [J,\exp(-iHs)] = 0 \, , \; \forall s \in {\mathbb R}, \quad
    \mbox{and} \quad \exp(-\beta H/2)\pi_\beta(A)\Omega_\beta =
    J\pi_\beta(A^*)\Omega_\beta \, , A \in {\mathcal A}.
    \]
  \end{enumerate}
\end{defi}
For quasifree states the definition can be reduced on the one-particle
Hilbert space. $K^{a\beta}_t$ maps to $\tilde{\mathcal H} = {\mathcal
  H} \oplus {\mathcal H}$, so the representation Hilbert space can be
chosen to be
\[
{\mathcal F} = {\mathcal F}_s({\mathcal H} \oplus {\mathcal H}) =
{\mathcal F}_s({\mathcal H}) \otimes {\mathcal F}_s({\mathcal H})  \, ,
\]
where ${\mathcal F}_s({\mathcal H})$ is the symmetric Fock space over
${\mathcal H}$. The Weyl operator $W(f)$ on ${\mathcal F}$ is
represented by
\[
W(f) = W_F(\cosh Z^\beta K^a_t f)\otimes W_F(C\sinh Z^\beta K^a_t f)
\, ,
\]
where $W_F$ is the usual Weyl operator on ${\mathcal F}_s$. With $h =
(m^2 - \Delta_\varepsilon/R^2)^{1/2}$ we define on $\tilde{\mathcal H}$
\[
  e^{-i\tilde{h}s} = e^{-ihs} \oplus e^{ihs}\, ,\quad
  e^{-\beta\tilde{h}/2} = e^{-\beta h/2} \oplus e^{\beta h/2}\, ,
\quad j(x\oplus y) = (-Cy) \oplus (-Cx) \, 
\]
and the operators on ${\mathcal F}$ by second quantization. 
The second condition in definition~\ref{kms} can be reduced to Kay's
regularity condition $KD \subset \mbox{dom}(h^{-1/2})$ (see Kay \cite{Kay85}).
$h$ is a positive, selfadjoint operator, so that $h^{-1/2}$ is
bounded and $\mbox{dom}(h^{-1/2}) = {\mathcal H}$. The condition $[J,e^{-iHs}] =
0$ reduces to $[j,e^{-i\tilde{h}s}] = 0$, which can easily be
verified. The condition $\exp(-\beta H/2)\pi_\beta(A)\Omega_\beta =
J\pi_\beta(A^*)\Omega_\beta$ reduces to $e^{-\beta \tilde{h}/2}(iK^{a\beta}f) =
j(-iK^{a\beta}f)$ and can also be verified. One finds $e^{-\beta
  \tilde{h}/2}(x\oplus y) = Cy\oplus Cx$, so that the one-particle KMS
condition 
\[
\skalar{e^{-is\tilde{h}}x}{y}_{\tilde{\mathcal H}} =
  \skalar{e^{-\beta\tilde{h}/2}y}{e^{-is\tilde{h}}
    e^{-\beta\tilde{h}/2}x}_{\tilde{\mathcal H}}   
\]
is valid for $x,y \in \tilde{{\mathcal H}}$ and all $s \in {\mathbb R}$.

\section{Time evolution by semigroups}
\label{sec:semigroups}

In this section we analyze the time evolution of adiabatic KMS states.
It is well known that the inverse temperature of a relativistic Bose
gas on Robertson-Walker spacetime is proportional to the scale
parameter $R$ of the metric, while for a non-relativistic gas the
inverse temperature is proportional to $R^2$. We will find the same
behavior for a Bose gas on a Robertson-Walker spacetime described by
an adiabatic KMS state.

We describe the time evolution on the classical phase space in terms
of semigroups and prove the existence of a propagator.

\subsection{The Klein-Gordon equation as a first order system}

To analyze the time evolution of an adiabatic KMS state, we first have
to describe the time evolution on the classical phase space $D$. This
can be achieved by introducing different coordinates, so that the
metric has the form
\[
g = -R^6(t)dt^2 + R^2(t)[d\theta_1^2 +
\Sigma^2_\varepsilon(d\theta_2^2 + \sin^2 \theta_2 d\phi^2]\, ,\quad
R(t)>0 \, , 
\]
where again $\varepsilon=-1,0,1$ corresponds to the spherical, the
flat and the hyperbolic spatial part respectively. We assume $R(t)$
and $\dot{R}(t)$ to be positive and continuous on any compact subset.
In these coordinates the Klein-Gordon equation has the form
\[   
[-\partial^2_t + R^4(t)\Delta_\varepsilon - m^2R^6(t)]\varphi = 0 \, ,
\]
where $\Delta_\varepsilon$ is again the Laplace operator on the
respective spatial spaces. The Klein-Gordon equation can be written as
a first order system:
\begin{eqnarray*}
\partial_t F &=& -H(t)F \\
F=\binom{f_1}{f_2}\,, \quad -H(t) &=&
\begin{pmatrix}
 0 & 1 \\ -B^2(t) & 0 
\end{pmatrix}
\,, \quad -B^2(t) = R^4(t)\Delta_\varepsilon - m^2R^6(t) \,,
\end{eqnarray*}
We define the operator $H(t)$ on the real Hilbert space ${\cal H}_t =
\mbox{dom}(B(t)) \oplus L^2(\EuS(t))$, where $\EuS$ are the respective
spatial spaces, with scalar product
\[
 \skalar{f_1 \oplus f_2}{g_1 \oplus g_2}_B = \skalar{Bf_1}{Bg_1} +
 \skalar{f_2}{g_2} \,.
\]
The phase space $D$ is of course a dense subspace of ${\mathcal H}_t$.
For fixed $s$ the operator $H(s)$ is skew-adjoint on this space and
therefore defines a contractive semigroup $T(t) = \exp [-tH(s)]$ on $\cal
H$. 

\subsection{Existence of the propagator}
\label{subsec:propagator}

We use the following theorem to prove the existence of the propagator.
For each positive integer $k$, we define an approximate propagator
$U_k(t,s)$ on $0 \leq s \leq t \leq 1$ by
\begin{eqnarray*}
U_k(t,s) &=&\exp\left( -(t-s)H\bigl((i-1)/k\bigr) \right) \\
\frac{i-1}{k} &\leq& 
s \leq t \leq \frac{i}{k} \,, \quad (\mbox{where} \quad 1 \leq i \leq
k) \\
 \mbox{and} \qquad \qquad
U_k(t,r) &=& U_k(t,s)U_k(s,r) \quad \mbox{if} \quad 0 \leq r
\leq s \leq t \leq 1 \,.
\end{eqnarray*}
We also define $C(t,s) = H(t)H(s)^{-1} - 1$.
\begin{theo}
Let $X$ be a Banach space and let $I$ be an open interval in $\mathbb
R$. For each $t \in I$, let $H(t)$ be the generator of a contraction
semigroup on $X$ so that $0 \in \rho(H(t))$, the resolvent set of
$H(t)$, and
\begin{enumerate}
  \item The $H(t)$ have a common dense domain $D$.
  \item For each $\varphi \in X$, $(t-s)^{-1}C(t,s)\varphi$ is uniformly
    strongly continuous and uniformly bounded in $s$ and $t$ for $t\neq
    s$ lying in any fixed compact subinterval of $I$.
  \item For each $\varphi \in X$, $C(t)\varphi \equiv \lim_{s \nearrow t}
    (t-s)^{-1}C(t,s)\varphi$ exists uniformly for $t$ in each compact
    subinterval and $C(t)$ is bounded and strongly continuous in $t$.
\end{enumerate}
Then for all $s\leq t$ in any compact subinterval of $I$ and any
$\varphi \in X$,
\[
  U(t,s)\varphi = \lim_{k\to \infty}U_k(t,s)\varphi
\]
exists uniformly in $s$ and $t$. Further, if $\psi \in D$, then
$\varphi_s(t) = U(t,s)\psi$ is in $D$ for all $t$ and satisfies
\[
  \frac{d}{dt}\varphi_s(t) = -H(t)\varphi_s(t) \, , \quad \varphi_s(s) = \psi
\]
and $\|\varphi_s(t)\| \leq \|\psi\|$ for all $t \geq s$.
\end{theo}
For a proof see \cite[Thm.~X.70]{ReedSimon2}.\\
As a consequence of the positivity of $B^2(t)$, $0$ is in the
resolvent set of $H(t)$. \\
We will verify condition 1, i.e.~we have to show for all $t \in I$ the
operators $H(t)$ have a common dense domain. For this it is sufficient
to show that the spaces $L^2(\EuS(t))$ are setwise equivalent. Let $h$
be the determinant of the spatial part of the metric $g$ and let
$\mu_h(t) = \sqrt{h(t)}$ and $\mu_h(t') = \sqrt{h(t')}$ be the
invariant measures induced by the metric $g$ on the Cauchy surfaces at
time $t$ and $t'$ respectively. Then
\[
\int_{\EuS} |f|^2 \sqrt{h(t)} \, d^3x =
\int_{\EuS} |f|^2\frac{\sqrt{h(t)}}{\sqrt{h(t')}}\sqrt{h(t')} \,d^3x \,,
\]
where $\sqrt{h(t)}/\sqrt{h(t')}$ is smooth, bounded and strictly
positive, namely the Radon-Nykodim derivative of the measure
$\mu_h(t)$ with respect to the measure $\mu_h(t')$. Therefore the
measures $\{\mu_h(t)\}_{t\in I}$ are mutually absolutely continuous
and because of the boundedness of $\sqrt{h(t)}/\sqrt{h(t')}$ we have
$f\in L^2(\EuS(t))$ iff $f\in L^2(\EuS(t'))$.\\
We have to verify condition 2. The operator $C(t,s)$ is given by
\begin{eqnarray*}
  C(t,s) &=& H(t)H(s)^{-1} - 1 = \\ 
  &=& \begin{pmatrix} 0 & -1 \\ B^2(t) & 0\end{pmatrix} 
  \begin{pmatrix} 0 & B^{-2}(s) \\ -1 & 0\end{pmatrix} - 
  \begin{pmatrix} 1 & 0 \\ 0 & 1\end{pmatrix} = 
  \begin{pmatrix} 0 & 0 \\ 0 & B^2(t)B^{-2}(s) - 1 \end{pmatrix} \,, 
\end{eqnarray*}
so we have
\[
  (t-s)^{-1}C(t,s)F = (t-s)^{-1}\left(\frac{m^2R^6(t) -
  R^4(t)\Delta_\varepsilon}{m^2R^6(s) - R^4(s)\Delta_\varepsilon} - 1
  \right)\pi \, , \quad F = \varphi \oplus \pi \in D \,.
\]
We assumed $R(t)$ to be continuous and as a consequence $R(t)$ is
jointly continuous on every compact subinterval. Therefore $C(t,s)$ is
jointly continuous in $t$ and $s$. The operator is also jointly
bounded on every compact subinterval (by using the eigenvalues of
$\Delta_\varepsilon$) because
$R(t)$ is bounded.\\
The last step is for $\varphi \in {\cal H}$ to show that $C(t)\varphi
= \lim_{s \nearrow t}(t-s)^{-1}(H(t)H(s)^{-1}-1)\varphi$ exists
uniformly for $t$ in each compact subinterval and $C(t)$ is bounded
and strongly continuous in $t$. The existence of the limit can be
shown with the rule of de l'Hospital. It is
\[
C(t) = 4\frac{\dot{R}(t)}{R(t)} +
2m^2\dot{R}(t)R(t)(m^2R^2(t)-\Delta_\varepsilon)^{-1} \,.
\]
The operator is bounded by the boundedness of
$(m^2R^2(t)-\Delta_\varepsilon)^{-1}$ and because we assumed $R(t)$
and $\dot{R}(t)$ to be bounded functions and it is of course strongly
continuous. \\
We have therefore proved the existence of the propagator $U(t,s)$ as
the strong limit of the approximate propagators $U_k(t,s)$.\\
{\bfseries Remark:} We have introduced new coordinates. In these
coordinates the existence proof is most easy. It is also possible to
proof the existence of the propagator using the coordinates given in
equation (\ref{metric}). Then the continuity and boundedness of
$\ddot{R}$ has to be assumed too.

\section{Time evolution on the one-particle Hilbert space}
\label{sec:time_evolution}

We will now describe the time evolution on the one-particle Hilbert
space. In definition~\ref{one_particle} we defined a one-particle
Hilbert space structure. For static spacetimes it is also required that
\begin{equation}\label{intertwine}
  U(t)K = K{\cal T}(t), 
\end{equation}
where ${\cal T}(t)$ describes the time evolution on $D$ and $U(t)$ the
time evolution on ${\cal H}$. i.e.~$K$ intertwines the time evolutions.

The time evolution on the classical phase space is given by the
propagator of subsection~\ref{subsec:propagator}. The propagator maps
Cauchy data at time $s$ to Cauchy data at time $t$. It is therefore
natural to require the following generalization of condition
(\ref{intertwine}).  For the one-particle Hilbert space structure
$K^a_t$ of an adiabatic vacuum state we demand that $K^a_t$
intertwines the time evolutions such that
\[
\tilde{U}(t,s)K^a_s = K^a_tU(t,s)\,.
\]
where $\tilde{U}(t,s)$ is time evolution on ${\cal H}$.  This is a
natural generalization. On the right hand side the evolution from a
Cauchy surface at time $s$ to a Cauchy surface at time $t$ is given on
phase space followed by the mapping to the one-particle Hilbert space
at time $t$. On the left hand side we map to the one-particle Hilbert
space at time $s$ and evolution is given on the Hilbert space to
a Cauchy surface at time $t$.

The operator $K_s^a$ is injective: For $F=f_1 \oplus f_2, G=g_1 
\oplus g_2 \in D$, we conclude from
\[
  \sqrt{2}K^a_s(F-G) = A^{1/4}(f_1-g_1) + iA^{-1/4}H(1+\frac{m^2}{2}A^{-1})
  (f_1-g_1) + iA^{-1/4}(f_2-g_2) = 0 \,,
\] 
that $f_1 = g_1$, because $A^{1/4}$ maps real-valued functions to
real-valued functions and this leads to $f_2 = g_2$, i.e.~$F = G$.
Since the kernel of $K_s^a$ contains only the zero vector, we can
define the propagator $\tilde{U}(t,s)$ on the one-particle Hilbert
space by
\[
  \tilde{U}(t,s) = K_t^aU(t,s)(K_s^a)^{-1}\,,
\]
on the range of $K_s^a$, which is dense in ${\mathcal H}$.

Furthermore we will show that $\tilde{U}(t,s)$ is isometric on the range of 
$K_s^a$ and can be extended to a unitary operator on ${\mathcal H}$.
The propagator $U(t,s)$ leaves the symplectic form $\sigma$ invariant,
because $\sigma$ is invariant under solutions of the Klein-Gordon
equation. Since the real-scalar product $S$ on $D$ can be defined by a
complexification $J$ ($J:D \to D$, $J^2 = -1$ see \cite[8.2.4]{BW}) via
\[
S(F,G) = \sigma(F,JG) \,,
\]
we also have $S(U(t,s)F,U(t,s)G) = S(F,G)$ and $U(t,s)J_s = J_tU(t,s)$,
where $J_s$ means the complexification, defining a state at time $s$.
Now with $f = K^a_sF, g = K^a_sG, F,G \in D$, we have 
\begin{eqnarray*}
  \skalar{f}{g} &=& \skalar{K_s^aF}{K^a_sG} = \\
  &=& [S(F,G) + i\sigma(F,G)]/2 = \\
  &=& [S(U(t,s)F,U(t,s)G) + i\sigma(U(t,s)F,U(t,s)G)]/2= \\
  &=&  [S(U(t,s)(K^a_s)^{-1}f,U(t,s)(K^a_s)^{-1}g) 
  + i\sigma(U(t,s)(K^a_s)^{-1}f,U(t,s)(K^a_s)^{-1}g)]/2= \\
  &=& \skalar{K^a_tU(t,s)(K^a_s)^{-1}f}{K^a_tU(t,s)(K^a_s)^{-1}g} =\\
  &=& \skalar{\tilde{U}(t,s)f}{\tilde{U}(t,s)g} \,,
\end{eqnarray*}
which shows the that $\tilde{U}(t,s)$ is an isometry and because
$\tilde{U}(t,s)$ is a bounded linear operator, densely defined on the
range of $K_s^a$, it can be extended to a unitary operator on
${\mathcal H}$.

The propagator $U(t,s)$ leaves the symplectic form $\sigma$ invariant.
Therefore it defines an automorphism $\alpha_{t,s}$ of the algebra
given by $\alpha_{t,s}(W(F)) = W(U(t,s)F)$. For an adiabatic vacuum
state $\omega$ we have
\begin{eqnarray*}
  \omega_t(\alpha_{t,s}(W(F))) &=& \omega_t(W(U(t,s)F)) =\\
  &=& \exp(-\|K^a_tU(t,s)F\|/4) = \\
  &=& \exp(-\|\tilde{U}(t,s)K^a_sF\|/4) =\\
  &=& \exp(-\|K^a_sF\|/4) = \\
  &=& \omega_s(W(F)) \, ,
\end{eqnarray*}
i.e.~$\omega_t \circ \alpha_{t,s} = \omega_s$.
  
\newpage
\section{Time evolution of an adiabatic KMS state}

In this section we will answer the question about the time evolution
of an adiabatic KMS state.

We start with the two-point function of an adiabatic KMS state of
inverse temperature $\beta_t$ on a Cauchy surface at time $t$ and
compute the two-point function on a Cauchy surface $s$.
\begin{eqnarray*}
  \lefteqn{ \langle K^{a\beta_t}_t (U(t,s)F) |K^{a\beta_t}_t
    (U(t,s)F)\rangle =}\\ &=& \left\langle K^a_t(U(t,s)F)
    \left|\frac{1+\exp[-\beta_t A^{1/2}(t)]}{1-\exp[-\beta_t
        A^{1/2}(t)]}K^a_t(U(t,s)F)\right. \right\rangle \\
  &=& \left\langle \tilde{U}(t,s)K^a_s(F) \left|\frac{1+\exp[-\beta_t
        A^{1/2}(t)]}{1-\exp[-\beta_t
        A^{1/2}(t)]}\tilde{U}(t,s)K^a_s(F)\right.\right\rangle \\
  &=& \left\langle K^a_s(F)\left|\frac{1+\exp[-\beta_t
        A^{1/2}(t)]}{1-\exp[-\beta_t
        A^{1/2}(t)]}K^a_s(F)\right.\right\rangle \,, \quad F \in D.
\end{eqnarray*}
We have 
\begin{eqnarray*}
  -\beta_t A^{1/2}(t) &=& -\beta_t \left(m^2 - \frac{\Delta}{R^2(t)}
  \right)^{1/2} \\ 
  &=& -\beta_t\frac{R(s)}{R(t)} \left(m^2\frac{R^2(t)}{R^2(s)} - 
    \frac{\Delta}{R^2(s)}\right)^{1/2}                    \\
  &=& -\beta_s\left(m^2\frac{R^2(t)}{R^2(s)} -
    \frac{\Delta}{R^2(s)}\right)^{1/2} \, , 
\end{eqnarray*}
so the state on the Cauchy surface $s$ can be interpreted as a state
of inverse temperature $\beta_s=\beta_t R(s)/R(t)$. This means that
the inverse temperature changes proportional to the scale parameter
$R$.

\section*{Appendix: Pseudodifferential operators and wave-front sets}

We shortly review in this appendix the necessary results on
pseudodifferential operators and wave-front sets (see e.g.~Taylor
\cite{Taylor}).

\begin{defi}
  Let $O$ be an open subset of ${\mathbb R}^n$.  We define the
  \textbf{symbol class} $S^m(O)$, $m \in {\mathbb R}$, to consist of
  all functions $p \in C^\infty(O \times {\mathbb R}^n)$ with the
  property that, for any compact set $K \subset O$, any multi-indices
  $\alpha,\beta$, there exists a constant $C_{K,\alpha,\beta}$ such
  that
  \begin{equation*}
    \left| D_x^\beta D_\xi^\alpha p(x,\xi) \right| \leq
    C_{K,\alpha,\beta} (1+|\xi|)^{m-|\alpha|}
  \end{equation*}
  for all $x \in K, \xi \in {\mathbb R}^n$.
\end{defi}
Remark: It is possible to define more general symbol classes but it is
not necessary for our work.\\
With each symbol $p(x,\xi)$ we associate the
\textbf{pseudodifferential operator} $P$ by
\[
  (Pf)(x) = (2\pi)^{-n} \int p(x,\xi)\widetilde{f}(\xi) e^{i\xi
  \cdot x}\,d\xi \, , \quad f\in {\mathcal S}({\mathbb R}^n)\, ,
\]
where $\;\widetilde{ }\;$ denotes the Fourier transform and if
$p(x,\xi) \in S^m$, we say $P \in OPS^m$. The operator $P$ is a
continuous operator of ${\mathcal D}({\mathbb R}^n)$ to
$C^\infty({\mathbb R}^n)$ and can be extended to a continuous operator
of ${\mathcal E}'({\mathbb R}^n)$ to ${\mathcal D}'({\mathbb R}^n)$.
By the Schwartz kernel theorem we can associate a distribution kernel
$K_P\in {\mathcal D}'({\mathbb R}^n \times {\mathbb R}^n)$ with the
map $P$ such that $\skalar{Pu}{v}=\skalar{K_P}{u\otimes v}$.
           
It is also possible to define a pseudodifferential operator (PDO) on a
paracompact manifold. 
\begin{defi}
  An operator $P:C_0^\infty(\M) \to C^\infty(\M)$ belongs to
  $OPS^m(\M)$ if the kernel of $P$ is smooth off the diagonal in $\M
  \times\M$ and for any coordinate neighborhood $U \subset \M$ there
  is a diffeomorphism $\chi:U \to O \subset \mathbb{R}^n$, such that
  the map of $C_0^\infty(O)$ into $C^\infty(O)$ given by $u \mapsto
  P(u\circ \chi)\circ \chi^{-1}$ belongs to $OPS^m(O)$.
\end{defi}
If $P \in OPS^m(O)$, we define the principal symbol of $P$ to be the
member of the equivalence class in $S^m(O)/S^{m-1}(O)$

Now we define the wave-front set of a distribution. If $p \in S^m(O)$
and $p_m$ its principal symbol, the characteristic set $char P$ of the
PDO $P$ associated with the symbol $p$ is given by
\[
  char P =  \{(x,\xi) \in T^*(O)\setminus \{0\} | p_m(x,\xi ) = 0\}\,.
\]
The \textbf{wave-front set} $WF(u)$ of a distribution $u$ is defined by
\[
  WF(u) = \bigcap \{ char P \, |\, P \in OPS^0,\; Pu \in C^\infty \} \,.
\]
A useful characterization of the wave-front set is given in the next
\begin{theo}
  The point $(x_0,\xi_0) \not\in WF(u)$ iff there is $\phi \in
  C^\infty_0(O)$, $\phi(x_0) \neq 0$, and a conic neighborhood
  $\Gamma$ of $\xi_0$, such that, for every $n$
  \begin{equation*}
    |\widetilde{\phi u} (\xi)| \leq C_n (1+|\xi|)^{-n}\, , \quad \xi
    \in \Gamma. 
  \end{equation*}
\end{theo}
For a proof see \cite[Chap.~VI \S 1]{Taylor}.
We quote two further results, important for this work.

\begin{theo}
\label{funktionalkalkuel}
Let $A \in OPS^1(\M)$ be an elliptic, selfadjoint, positive operator
on a compact manifold $\M$ with real valued principal symbol
$a(x,\xi)$.  Let $p(\lambda) \in S^m({\mathbb R})$ be a Borel
function. Then $p(A) \in OPS^m(\M)$ with principal symbol
$p(a(x,\xi))$.
\end{theo}
For a proof see \cite[Chap.~XII \S 1]{Taylor}.\\
We remark that the square-root of the Laplace operator
on a compact manifold is of this type, $(-\Delta)^{1/2} \in OPS^1(\M)$
with principal symbol given by $\sqrt{h^{ij}\xi_i\xi_j}$, where
$h_{ij}$ are the metric coefficients.
\begin{theo}
\label{glattheit}
  \begin{enumerate}
  \item If $A \in OPS^m(O)$, then the associated kernel distribution
    $K_A$ is smooth everywhere off the diagonal in $O \times O$.
  \item If $A \in OPS^{-\infty}(O)$, then $K_A$ is smooth everywhere
    in $O \times O$.
  \end{enumerate}
\end{theo}
For a proof see \cite[Lemma 2.6]{Junker}.\\[1cm]
{\bf Acknowledgments.} I thank K.-E.~Hellwig, W.~Junker and M.~Keyl for
helpful hints and discussions.

\end{document}